\documentclass{cimento}

\usepackage{graphicx}  

\title{Properties of pasta phases in catalyzed neutron stars}
\author{H.~Dinh Thi\from{ins:x}\ETC,
A. F.~Fantina\from{ins:y},
\atque
F.~Gulminelli\from{ins:x}
        }
\instlist{\inst{ins:x} Laboratoire de Physique Corpusculaire (LPC), CNRS, ENSICAEN, UMR6534, Université de Caen Normandie, F-14050 Caen Cedex, France
  \inst{ins:y} Grand Accélérateur National d'Ions Lourds (GANIL), CEA/DRF - CNRS/IN2P3, Boulevard Henri Becquerel, 14076 Caen, France}
\begin{document}

\maketitle

\begin{abstract}
Exotic non-spherical configurations of nuclei, known as ``pasta'' phases,  are expected to be present at the bottom of the inner crust of a neutron star. 
We study the properties of these configurations in catalyzed neutron stars within a compressible liquid-drop model approach, with surface parameters optimized to reproduce experimental nuclear masses.
Our results show that the properties of the pasta phases exhibit strong model dependence. 
To estimate the model uncertainties, a Bayesian analysis is performed, combining information from nuclear physics experiments and chiral perturbation theoretical calculations with astrophysical observations. 
The inferred posterior distributions are discussed, with particular focus on the effect of the low-density energy functional on the predictions.
\end{abstract}

\section{Introduction}
\label{sec:intro}
In the deepest region of the inner crust  of a neutron star (NS), it is expected that non-spherical shapes of nuclei, known as ``pasta phases'',  are energetically favorable. 
The existence of pasta phases, in particular of ``bubbles'', at high densities in the crust was already suggested in the pioneering work of ref.~\cite{ref:bpp}. 
Other, non-spherical, shapes, particularly two-dimensional (rods, tubes) and one-dimensional (slabs) geometries, were further investigated in refs. \cite{ref:Ravenhall1983, ref:Hashimoto1984, ref:oya1984}, while more complex structures were studied more recently, e.g., in refs.~\cite{ref:gyroid1, ref:gyroid2, ref:waffles, ref:Alcain}.

Numerous studies have been conducted in the last decades on the nuclear pasta, using different approaches, like compressible liquid drop models (CLDMs), semi-classical methods, nuclear energy density functional theory, and molecular dynamics (see, e.g., ref.~\cite{ref:blacha2018} for a recent review, and references therein).
Recently, we have extended the CLDM of Carreau \textit{et al.} \cite{ref:Carreau2019, ref:Carreau2020} to account for non-spherical configurations in the NS inner crust \cite{ref:HoaAA21, ref:HoaEPJA21}, and evaluated the uncertainties in the pasta-phase properties using the Bayesian inference. 
This analysis predicted that a sizable amount of the crust is made up of pasta structures. 
Specifically, we estimated the relative thickness and mass of the pasta layer compared to that of the whole crust to be $12.8\%$ and $48.5\%$, respectively (see table 5 in ref.~\cite{ref:HoaAA21}), in good agreement with the results obtained in ref.~\cite{ref:balliet21}. 
Although, up to now, there is no direct evidence of the existence of the pasta phases from NS observations, the presence of these ``exotic" configurations can potentially have an important impact on various NS phenomena, such as NS cooling \cite{ref:cool1, ref:cool2, ref:cool3}, transport properties \cite{ref:trans1, ref:trans2}, crust oscillations \cite{ref:os1, ref:os2}, and pulsar magnetic and rotational evolution \cite{ref:pons13}.  

In this work, we pursue the investigation of refs.~\cite{ref:HoaAA21, ref:HoaEPJA21}, presenting additional discussions on both the model dependence of the pasta-phase predictions and the effect of the low-energy density functional on the fractional crustal thickness and mass of the pasta layer.
In sect.~\ref{sec:formalism}, we describe the main points of the formalism used in modeling the nuclear pasta.
The model dependence of the pasta-phase properties is presented in sect.~\ref{sec:model_dependence}, while in sect.~\ref{sec:bayes} we discuss the importance of constraining the energy functional at low densities for the determination of the uncertainties of pasta-phase observables. 
Finally, conclusions are given in sect.~\ref{sec:conclusion}.

\section{Model of the pasta phases}
\label{sec:formalism}
We model the pasta phases under the cold catalyzed matter hypothesis, that is, under the assumption of full equilibrium at zero temperature. 
Details on the formalism are presented in refs.~\cite{ref:HoaAA21, ref:HoaEPJA21}; here, we briefly recall the main points and assumptions.

At a given baryon density $n_B$ in the NS inner crust, the structure of matter is approximated by a periodic lattice configuration of  identical Wigner-Seitz (WS) cells  of volume $V_{\rm WS}$. 
Each WS cell contains a cluster or a hole, of volume $V$. 
The density distribution in the Wigner–Seitz cell is $n_i$ ($n_g$) if $l < r_N$, and $n_g$ ($n_i$) otherwise, $n_g$ being the density of the surrounding uniform neutron gas, $r_N$ is the linear dimension of the pasta structure, and $l$ the linear coordinate of the cell. 
The density of the denser phase is $n_i = A/V$ in the case of cluster ($n_i = A/(V_{\rm WS} - V)$ in the case of holes) and its proton fraction is given by $y_p = Z/A$, $Z$ ($A$) being the proton number (total mass number) of the cluster.
The surrounding uniform electron gas has density $n_e$ such that charge neutrality holds, \textit{i.e.}, $n_e = n_p$, with $n_p$ being the total proton density in the cell.

In order to obtain the ground state of the system, the energy density of the WS cell has to be minimized  with the constraint of baryon density conservation. The corresponding thermodynamic potential per unit volume in the CLDM can be written as: 
\begin{eqnarray}
\Omega
&=& (n_B-n_p) m_n c^2+n_p m_p c^2 + \epsilon_B(n_i,1-2y_p)f(u) \nonumber \\ 
&+& \epsilon_B(n_g,1) (1-f(u)) + \epsilon_{\rm Coul} + \epsilon_{\rm {surf+curv}} + \epsilon_e(n_e) -\mu_B^{\rm tot} \ n_B , 
\label{eq:auxiliary}
\end{eqnarray}
where $m_n$ ($m_p$) is the neutron (proton) mass, $\epsilon_B(n,\delta)$ is the uniform nuclear matter energy density at density $n$ and isospin asymmetry $\delta=(n_n-n_p)/n$, with $n_n$ ($n_p$) the neutron (proton) density, $\epsilon_e (n_e)$ is the electron gas energy density, $\mu_B^{\rm tot}$ is the baryonic chemical potential (including the rest mass), $\epsilon_{\rm {surf+curv}}$ and $\epsilon_{\rm Coul}$ are the surface and Coulomb energies, respectively, and the function $f$ is given by $f(u) = u$ (or $f(u)=1-u$) for clusters (holes), with $u=V/V_{\rm WS}$; see sect.~2 in ref.~\cite{ref:HoaEPJA21} for details.

From eq.~(\ref{eq:auxiliary}), we can see that a nuclear model is defined by the choice of the bulk functional, $\epsilon_B$, supplemented with the interface energy, $\epsilon_{\rm {surf+curv}}$ and $\epsilon_{\rm Coul}$. 
For the energy density of homogeneous nuclear matter, $\epsilon_B$, we use the meta-modeling approach proposed by Margueron \textit{et al.} \cite{ref:metamodel1, ref:metamodel2}. 
Within this approach, a Taylor expansion in $x=(n-n_{\rm sat})/(3n_{\rm sat})$ ($n_{\rm sat} \approx 0.15$~fm$^{-3}$ being the saturation density) up to order $N$ ($N=4$ in the present case) around the saturation point ($n=n_{\rm sat}$, $\delta=0$) is introduced, with the parameters of the expansion corresponding to the so-called equation-of-state empirical parameters. 
In addition, a $\delta^{5/3}$ term from the fermionic zero-point energy and an exponential correction ensuring the correct limiting behavior at zero density are added (see eq.~(5) in ref.~\cite{ref:HoaEPJA21}, and ref.~\cite{ref:Carreau2019} for details). 
As for the finite-size contributions, which are the only terms dependent on the nuclear geometry, we make use of the same expressions as in refs.~\cite{ ref:HoaAA21, ref:HoaEPJA21, ref:newton13}, see eqs.~(8)-(15) in ref.~\cite{ref:HoaEPJA21}. 
 
The complete parameter set, denoted as $\bf X$, consists of 18 parameters (13 bulk parameters plus 5 surface parameters): for each functional, i.e. for a given set of bulk parameters, the surface and curvature parameters are determined from a $\chi^2$-fit to the experimental masses from the Atomic Mass Evaluation (AME) 2016 \cite{ref:ame2016}.
Then, given a nuclear model, i.e. a full set of $\bf{X}$ parameters, at each baryon density $n_B$, the equilibrium phase and composition are obtained in two steps: first, the optimal geometry is determined by the configuration having the lowest WS-cell energy density, then, the thermodynamic potential, eq.~(\ref{eq:auxiliary}), is minimized with respect to five variational parameters ($n_i$, $I=1-2y_p$, $A$, $n_p$, and $n_g$) to obtain the composition.
Note that we consider here five customary geometries, namely spheres, cylinders (rods), slabs, tubes, and bubbles.
The equilibrium phase corresponds to the one having the minimum value of the ``optimal'' thermodynamic potential.

\section{Model dependence of the pasta phases}
\label{sec:model_dependence}

To show the model dependence of the pasta-phase properties, we performed the calculations as described in sect. \ref{sec:formalism} employing three different nuclear energy functionals, as illustrative examples: BSk24 \cite{ref:bsk24}, DD-ME$\delta$ \cite{ref:ddmed}, and TM1 \cite{ref:tm1}. 
The corresponding values of empirical parameters and optimized surface parameters for these three functionals are given in tables 1 and 2 of ref.~\cite{ref:HoaAA21}.

\begin{figure}[htbp]
\centering
\includegraphics[scale = 0.33]{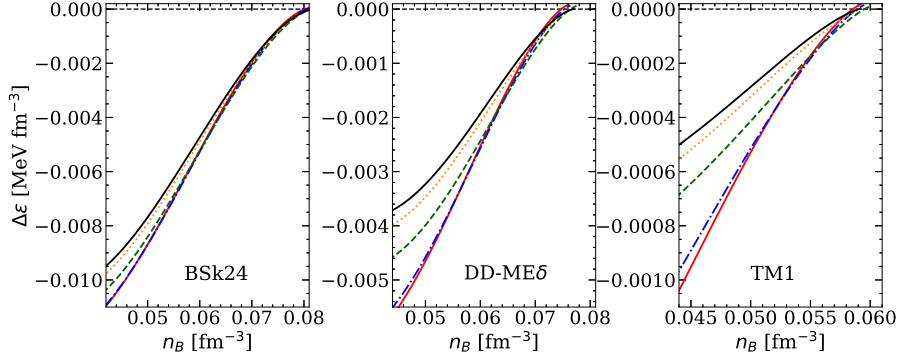}     
\caption{WS-cell energy-density difference as a function of baryon density for five geometries: spheres (red solid lines), rods (blue dash-dotted lines), slabs (green dashed lines), tubes (orange dotted lines), and bubbles (black solid lines), for three functionals: BSk24, DD-ME$\delta$, and TM1. In each panel, the point of intersection of the lowest $\Delta \epsilon$ curve with the horizontal black line corresponds to the crust-core transition.}
\label{fig:delta_epsilon}
\end{figure}

\begin{figure}[htbp]
\centering
\includegraphics[scale = 0.35]{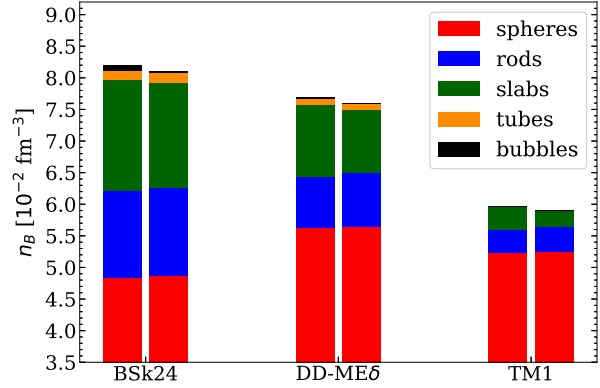}      
\caption{Sequence of equilibrium geometries as a function of baryon density $n_B$ for three nuclear functionals: BSk24, DD-ME$\delta$, and TM1. See text for details.
}
\label{fig:equi}
\end{figure}

We start the discussion by showing in fig.~\ref{fig:delta_epsilon} the competition among the geometries in the densest region of the inner crust, near the crust-core transition. 
Since, as mentioned in sect.~\ref{sec:formalism}, the only dependence on the geometry of the pasta in the thermodynamic potential to be minimized, eq.~(\ref{eq:auxiliary}), enters in the surface and Coulomb energies, in fig.~\ref{fig:delta_epsilon} we plot the energy difference $\Delta \epsilon = \epsilon_{\rm crust} - \epsilon_{HM}$, where $\epsilon_{\rm crust}$ is the WS-cell energy density obtained from the minimization procedure for a given geometry, and $\epsilon_{\rm HM}$ is the WS-cell energy density calculated for the homogeneous matter at $\beta$-equilibrium. 
We can clearly observe that the differences in $\Delta \epsilon$ are more pronounced when comparing the three functionals than among the five geometries within the same functional. 
This also yields, for the three models, very different optimal compositions, as one can see, for example, in fig.~2 of ref.~\cite{ref:HoaEPJA21}. 
In addition, the results in fig.~\ref{fig:delta_epsilon} show that the transitions among the five phases as well as the crust-core transition are strongly affected by the choice of the energy functional. 
This can  also be observed comparing the left columns illustrated for each functional in fig. \ref{fig:equi}, where the sequence of equilibrium geometries is displayed.

\begin{figure}[htbp]
\centering
\includegraphics[scale = 0.37]{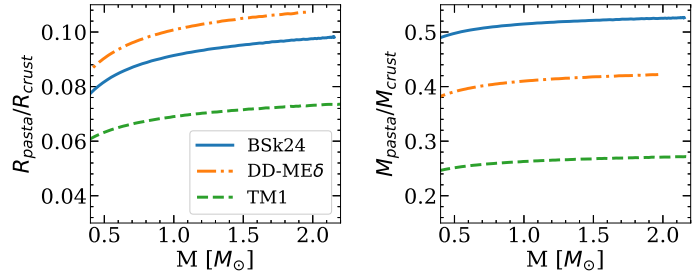}      
\caption{Thickness (left panel) and mass (right panel) fractions of the pasta phases with respect to the whole crust as a function of the NS mass $M$ for three functionals: BSk24 (blue solid lines), DD-ME$\delta$ (orange dash-dotted lines), and TM1 (green dashed lines).}
\label{fig:obs}
\end{figure}

The consistent minimization of the thermodynamic potential $\Omega$ in eq.~(\ref{eq:auxiliary}) for each geometry is crucial in getting the correct configuration for the ground state of matter. 
This procedure, which is however relatively time consuming, plays an important role in calculations requiring the exact composition as input, such as in the calculations of transport coefficients \cite{ref:trans2}. 
However, for the determination of more global quantities, such as thickness, mass, or moment of inertia of the pasta phases, one can instead fix the composition of the different phases to that obtained for spheres.
To justify this point, on the right columns for each functional in fig.~\ref{fig:equi}, we display the sequence of the equilibrium geometries obtained when the minimization is performed assuming for all geometries the composition obtained for spheres.
The results suggest that this assumption does not change considerably the transition densities, and can thus be adopted to reduce the computational time when performing the Bayesian analysis (see next section).

A complementary information on the importance of the pasta layer in NSs is given by the pasta fractional thickness and the associated mass, that we plot  in fig.~\ref{fig:obs} as a function of the NS mass $M$ (in unit of solar mass $M_{\odot}$). Although the numerical values of $R_{\rm pasta}/R_{\rm crust}$ and $M_{\rm pasta}/M_{\rm crust}$ are strongly model dependent, the presence of pasta is robustly predicted by all the considered models. 
On the other hand, the variation of the fractional pasta thickness (mass) with the NS mass is similar for the three functionals and amounts to $\approx 20\%$ ($\approx 10\%$) when changing the NS mass from $0.4 M_\odot$ to $2 M_\odot$.

\section{Influence of the low-density equation of state}
\label{sec:bayes}

The results presented in the previous section highlight the model dependence of pasta properties. 
To quantitatively assess this point and determine the impact of our current incomplete knowledge of the nuclear energy functional on the uncertainties of pasta-phase properties, we have performed a Bayesian analysis.
We started by considering flat non-informative priors, constructed by largely varying the model parameters $\bf {X}$ according to our current knowledge provided by nuclear-physics experiments (see table 1 in ref. \cite{ref:HoaEPJA21}). 
For each prior model, the minimization of the corresponding $\Omega$ potential, eq.~(\ref{eq:auxiliary}), was carried out to find the equilibrium composition of the crust. filter
Models yielding non-physical solutions were discarded.
We then applied to the prior both low-density (LD) constraints from nuclear physics and high-density (HD) constraints coming from general and NS physics to generate the posterior distribution:
\begin{equation}
p_{\rm post} ({\bf{X}}) = \mathcal{N} \, \omega_{\rm LD}({\bf{X}}) \, \omega_{\rm HD}({\bf{X}}) \, \omega_{\rm mass} \, p_{\rm prior}({\bf{X}})  ,
\label{eq:probalikely}
\end{equation}
where $ \mathcal{N}$ is the normalization factor, $p_{\rm prior}$ is the prior probability, and $\omega_{\rm mass}$ is a likelihood expression representing the quality of reproduction of the experimental masses from the AME2016 \cite{ref:ame2016}.
The  $w_{\rm LD}$  filter accounts for the uncertainty band of the chiral effective field theory (EFT) calculations of the energy per particle of symmetric and pure neutron matter of ref.~\cite{ref:drischler16}, which is considered as a $90\%$ confidence interval.
Due to the accuracy in the chiral EFT calculation, the filtering bands at low density are very narrow, thus the rejection rate is high. 
The second strict filter we applied, $\omega_{\rm HD}$, is defined by imposing causality, thermodynamic stability, non-negative symmetry energy, and the resulting equation of state to support observed massive NSs \cite{ref:antoniadis13}, $M_{\rm max} \geq 1.97 M_{\odot}$, where $M_{\rm max}$ is the maximum NS mass obtained with a given model, i.e. with a given parameter set $\bf X$ (see refs.~\cite{ref:gulfan21, ref:HoaUniv21} for a discussion on the importance of the high-density constraints).

\begin{figure}[htbp]
    \centering
    \includegraphics[scale = 0.4]{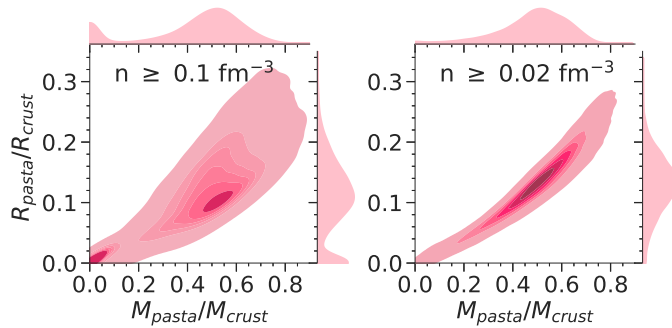}
    \label{fig:bayes}
    \caption{Posterior correlation between the fractional thickness and mass of the pasta phases with respect to the whole crust calculated at $M = M_{\rm max}$. Left (right) panel: the LD filter is applied in the interval $\left[0.1, 0.2\right]$ fm$^{-3}$ ($\left[0.02, 0.2\right]$ fm$^{-3}$). }
\end{figure}

Since NS global properties are rather determined by the high-density part of the equation of state, the compatibility of the functionals with ab-initio predictions at very low-density was usually overlooked.
Here, with the aim of focusing on the importance of the low-density part of the energy functional on the pasta-phase properties, we considered two density intervals for applying the LD filter, namely $\left[0.1, 0.2\right]$ fm$^{-3}$ and $\left[0.02, 0.2\right]$ fm$^{-3}$. 
When the filter was applied in the range $\left[0.1, 0.2\right]$ ($\left[0.02, 0.2\right]$) fm$^{-3}$, we generated $2\times 10^6$ ($10^8$) models in the prior, and obtained 7714 (7008) models in the final posterior. 
These statistics are sufficient for this study as increasing the number of models has no significant impact on the results. 
The posterior correlation of the fractions of thickness, $R_{\rm pasta}/R_{\rm crust}$, and associated mass, $M_{\rm pasta}/M_{\rm crust}$, as well as their probability density distributions, are displayed in fig. \ref{fig:bayes}. 
When the filter is applied from $n\geq 0.1$ fm$^{-3}$ (left panel), noticeable peaks arise at very low $R_{\rm pasta}/R_{\rm crust}$ and $M_{\rm pasta}/M_{\rm crust}$, compatible with a small or even null pasta layer, contrarily to the case when the energy functional is constrained from lower density, $n\geq 0.02$ fm$^{-3}$ (right panel). 
In the latter case, the uncertainties in the fractional crustal thickness and mass of the pasta layer are reduced and the correlation enhanced. 
These results corroborate those obtained in refs.~\cite{ref:HoaAA21,ref:HoaEPJA21}, additionally showing the importance of constraining the energy functional in the very-low-density region to determine the pasta-phase properties (see refs.~\cite{ref:HoaAA21, ref:HoaEPJA21} for a discussion on the correlation of the pasta-phase properties with the nuclear parameters $\bf{X}$).

\section{Conclusions}
\label{sec:conclusion}

To study the pasta-phase properties in catalyzed NSs, we employed a CLDM in which the bulk energy is calculated using a meta-modeling approach and finite-size parameters are optimized on the AME2016 atomic mass table.
The model dependence of the results is illustrated using three energy functionals, namely BSk24, DD-ME$\delta$, and TM1.
To quantitatively estimate the uncertainties of the pasta-phase properties, we performed a Bayesian analysis starting from non-informative priors obtained largely varying the model parameters within current uncertainties coming from nuclear-physics experiments and imposing constrains coming from both (theoretical and experimental) nuclear physics and astrophysics. 
The results we obtained confirm those presented in our previous studies \cite{ref:HoaAA21, ref:HoaEPJA21}, namely, (i) the robustness in the prediction of the existence of pasta phases at the bottom of the inner crust; (ii) the model dependence of the pasta-phase properties; and (iii) the importance of constraining the functional from very low densities to properly determine the pasta-phase observables.

\acknowledgments
The authors acknowledge partial support from the IN2P3 Master Project ``NewMAC'' and the CNRS International Research Project (IRP) ``Origine des éléments lourds dans l'univers: Astres Compacts et Nucléosynthèse (ACNu)''.


\begin{thebibliography}{0}

\bibitem{ref:bpp} \BY{Baym G., Bethe H. A. \atque Pethick C. J.} \IN{Nucl. Phys. A}{175}{1971}{225-271}.

\bibitem{ref:Ravenhall1983} \BY{Ravenhall D. G., Pethick C. J. \atque Wilson J. R.} \IN{Phys. Rev. Lett.}{50}{1983}{2066-2069}.

\bibitem{ref:Hashimoto1984} \BY{Hashimoto M., Seki H. \atque  Yamada M.} \IN{Prog. Theor. Phys.}{71}{1984}{320-326}.

\bibitem{ref:oya1984} \BY{Oyamatsu K.,  Hashimoto M. \atque  Yamada M.} \IN{Prog. Theor. Phys.}{72}{1984}{373}.

\bibitem{ref:gyroid1} \BY{Nakazato K., Oyamatsu K. \atque Yamada S.} \IN{Phys. Rev. Lett.}{103}{2009}{132501}.

\bibitem{ref:gyroid2} \BY{Schuetrumpf B., Klatt M. A., Iida K., et al.} \IN{Phys. Rev. C}{91}{2015}{025801}.

\bibitem{ref:waffles} \BY{Schneider A. S., Berry D. K., Briggs C. M., Caplan M. E. \atque Horowitz C. J.} \IN{Phys. Rev. C}{90}{2014}{055805}.

\bibitem{ref:Alcain} \BY{Alcain P. N., Giménez Molinelli P. A., \atque Dorso C. O.} \IN{Phys. Rev. C}{90}{2014}{065803}.

\bibitem{ref:blacha2018} \BY{Blaschke, D. \atque Chamel, N.} in \TITLE{The Physics and Astrophysics of Neutron Stars}, vol. 457, edited by \NAME{Rezzolla L.,  Pizzochero P.,  Jones D.I.,  Rea N. \atque  Vidaña I.} (Astrophysics and Space Science Library, Springer, Cham) 2018, pp.~337-400.

\bibitem{ref:Carreau2019} \BY{Carreau T.,  Gulminelli F. \atque  Margueron J.} \IN{Eur. Phys. J. A}{55}{2019}{188}.

\bibitem{ref:Carreau2020} \BY{Carreau T.,  Gulminelli F.,  Chamel N.,  Fantina A. F. \atque  Pearson J. M.} \IN{Astron. Astrophys.}{635}{2020}{A84}.

\bibitem{ref:HoaAA21} \BY{Dinh Thi H.,  Carreau T., Fantina A. F. \atque Gulminelli F.} \IN{Astron. Astrophys.}{654}{2021}{A114}.

\bibitem{ref:HoaEPJA21} \BY{Dinh Thi H., Fantina A. F. \atque Gulminelli F.} \IN{Eur. Phys. J. A}{57}{2021}{296}.

\bibitem{ref:balliet21} \BY{Balliet L. E., Newton W. G., Cantu S. \atque Budimir S.} \IN{Astrophys. J.}{918}{2021}{79}.

\bibitem{ref:cool1} \BY{Newton W. G.,  Murphy K.,  Hooker J. \atque Li B. A.} \IN{Astrophys. J. Lett.}{779}{2013}{L4}.

\bibitem{ref:cool2} \BY{Horowitz C. J.,  Berry D. K.,  Briggs C. M., et al.} \IN{Phys. Rev. Lett.}{114}{2015}{031102}.

\bibitem{ref:cool3} \BY{Lin Z.,  Caplan M. E. , Horowitz C. J. \atque Lunardini C.} \IN{Phys. Rev. C}{102}{2020}{045801}.

\bibitem{ref:trans1} \BY{Yakovlev D. G.} \IN{Mon. Not. R. Astron. Soc.}{453}{2015}{ 581–590}.

\bibitem{ref:trans2} \BY{Schmitt A. \atque  Shternin P.} in \TITLE{The Physics and Astrophysics of Neutron Stars}, vol. 457, edited by \NAME{Rezzolla L.,  Pizzochero P.,  Jones D.I.,  Rea N. \atque  Vidaña I.} (Astrophysics and Space Science Library, Springer, Cham) 2018,  pp.~455–574.

\bibitem{ref:os1}  \BY{Gearheart M.,  Newton W. G.,  Hooker J. \atque  Li B. A.} \IN{Mon. Not. R. Astron. Soc.}{418}{2011}{2343-2349}.

\bibitem{ref:os2} \BY{Sotani H.,  Nakazato K.,  Iida K. \atque  Oyamatsu K.} \IN{Phys. Rev. Lett}{108}{2012}{201101}.

\bibitem{ref:pons13} \BY{Pons J.A.,  Viganò D. \atque  Rea N.} \IN{Nat. Phys.}{9}{2013}{431–434}.

\bibitem{ref:metamodel1} \BY{Margueron J.,  Hoffmann Casali R. \atque  Gulminelli F.} \IN{Phys. Rev. C}{97}{2018}{025805}.

\bibitem{ref:metamodel2} \BY{Margueron J.,  Hoffmann Casali R. \atque  Gulminelli F.} \IN{Phys. Rev. C}{97}{2018}{025806}.


\bibitem{ref:newton13} \BY{Newton W.G.,  Gearheart M. \atque Li B.A.} \IN{Astrophys. J. Suppl. Ser.}{204}{2013}{9}.

\bibitem{ref:ame2016} \BY{Wang M.,  Audi G.,  Kondev F. G., Huang  W.J.,  et al.} \IN{Chin. Phys. C}{41}{2017}{030003}.  

\bibitem{ref:bsk24} \BY{Goriely S., Chamel N., \atque Pearson J. M.} \IN{Phys. Rev. C}{88}{2013}{024308}.

\bibitem{ref:ddmed} \BY{Roca-Maza X., Viñas X., Centelles M., Ring P. \atque Schuck P.} \IN{Phys. Rev. C}{84}{2011}{054309}.

\bibitem{ref:tm1} \BY{Shen H., Toki H., Oyamatsu K. \atque Sumiyoshi K.} \IN{Nucl. Phys. A}{637}{1998}{435}.

\bibitem{ref:drischler16} \BY{Drischler C., Hebeler K. \atque Schwenk A.} \IN{Phys. Rev. C}{93}{2016}{054314}.

\bibitem {ref:antoniadis13} \BY{Antoniadis J., Freire P. C. C., Wex N., et al.} \IN{Science}{340}{2013}{6131}.

\bibitem{ref:gulfan21} \BY{Gulminelli, F. \atque Fantina, A.~F.} \IN{Nucl. Phys. News}{31}{2021}{2}.

\bibitem{ref:HoaUniv21} \BY{Dinh Thi, H., Mondal, C. \atque Gulminelli, F.} \IN{Universe}{7}{2021}{373}.






\end{thebibliography}
\end{document}